\documentclass[usenames,dvipsnames]{acm_modified}
\usepackage{caption} 
\usepackage{booktabs}
\usepackage{enumitem} 
\usepackage{wrapfig} 
\usepackage{tcolorbox} 
\usepackage{tabularx}
\usepackage{array}
\usepackage{colortbl}
\usepackage{soul}
\tcbuselibrary{skins}
\newcolumntype{Y}{>{\raggedleft\arraybackslash}X}
\tcbset{tab1/.style={fonttitle=\bfseries\large,fontupper=\normalsize\sffamily,
colback=yellow!10!white,colframe=red!75!black,colbacktitle=Salmon!40!white,
coltitle=black,center title,freelance,frame code={
\foreach \n in {north east,north west,south east,south west}
{\path [fill=red!75!black] (interior.\n) circle (3mm); };},}}
\tcbset{tab2/.style={enhanced,fonttitle=\bfseries,fontupper=\normalsize\sffamily,
colback=yellow!10!white,colframe=red!50!black,colbacktitle=Salmon!40!white,
coltitle=black,center title}}
\newcolumntype{x}{>{\centering\arraybackslash}X}
\newcolumntype{s}{>{\centering\arraybackslash}>{\hsize=.25\hsize}X}

\usepackage{adjustbox}
\usepackage{floatrow} 

\newcommand{\ds}{\displaystyle}
\newcommand{\df}{\displaystyle\frac}

\newcommand{\Chi}{\chi}
\newcommand{\quotes}[1]{``#1''}

\newcommand{\ra}[1]{\renewcommand{\arraystretch}{#1}}

\newcommand*\justify{%
  \fontdimen2\font=0.4em
  \fontdimen3\font=0.2em
  \fontdimen4\font=0.1em
  \fontdimen7\font=0.1em
  \hyphenchar\font=`\-
}

\usepackage{lmodern}
\makeatletter
  \newcommand{\miniscule}{\@setfontsize\miniscule{4.5}{5.5}}
\makeatother

\begin{document} \sloppy











\title{Wavelet decomposition of software entropy \\
reveals symptoms of malicious code}

%
%
%
%
%

\numberofauthors{1} 
%
\author{
%
%
\alignauthor Michael Wojnowicz, Glenn Chisholm, Matt Wolff, Xuan Zhao \\
	\affaddr{Dept. of Research and Intelligence} \\
        \affaddr{Cylance, Inc.}\\
       \affaddr{18201 Von Karman Drive}\\
       \affaddr{Irvine, CA 92612}\\
       \email{ \{mwojnowicz, gchisholm, mwolff, xzhao\} @cylance.com} 
}


\maketitle

\begin{abstract}

Sophisticated malware authors can sneak hidden malicious contents into portable executable files, and these contents can be hard to detect, especially if encrypted or compressed.  However, when an executable file switches between content regimes (e.g., native, encrypted, compressed, text, and padding), there are corresponding shifts in the file's representation as an entropy signal.  In this paper, we develop a method for automatically quantifying the extent to which patterned variations in a file's entropy signal make it ``suspicious."  In Experiment 1, we use wavelet transforms to define a Suspiciously Structured Entropic Change Score (SSECS), a scalar feature that quantifies the suspiciousness of a file based on its distribution of entropic energy across multiple levels of spatial resolution.  Based on this single feature, it was possible to raise predictive accuracy on a malware detection task from 50.0\% to 68.7\%, even though the single feature was applied to a heterogeneous corpus of malware discovered ``in the wild."   In Experiment 2, we describe how wavelet-based decompositions of software entropy can be applied to a parasitic malware detection task involving large numbers of samples and features.  By extracting only string and entropy features (with wavelet decompositions) from software samples, we are able to obtain almost 99\% detection of parasitic malware with fewer than 1\% false positives on good files.   Moreover, the addition of wavelet-based features uniformly improved detection performance across plausible false positive rates, both in a strings-only model (e.g., from 80.90\% to 82.97\%) and a strings-plus-entropy model (e.g. from 92.10\% to 94.74\%, and from 98.63\% to 98.90\%). Overall, wavelet decomposition of software entropy can be useful for machine learning models for detecting malware based on extracting millions of features from executable files.\footnote{This article is a post-print of~\cite{wojnowicz4} which corrects typos introduced during editing.}

\end{abstract}

\section{Introduction}
\subsection{The Entropy Of Malicious Software}

A fundamental goal in the information security industry is malware detection. In this paper, we focus our malware detection efforts on the fact that malicious files (e.g. parasitics, or exploits with injected shellcode) commonly contain encrypted or compressed (\quotes{packed}) segments which conceal malicious contents  \cite{BM}.   Thus, the information security industry has been interested in developing methodologies which can automatically detect the presence of encrypted or compressed segments hidden within portable executable files.   To this end, entropy analysis has been used, because files with high entropy are relatively likely to have encrypted or compressed sections inside them \cite{LH}.    In general, the entropy of a random variable reflects the amount of uncertainty (or lack of knowledge) about that variable.   In the context of software analysis, zero entropy  would mean that the same character was repeated over and over (as might occur in a \quotes{padded} chunk of code), and maximum entropy would mean that a chunk consisted of entirely distinct values.    Thus, chunks of code that have been compressed or encrypted tend to have higher entropy than native code.   For instance, in the software corpus studied by~\cite{LH}, plain text had an average entropy of 4.34, native executables had an average entropy of 5.09, packed executables had an average entropy of 6.80, and encrypted executables had an average entropy of 7.17. 


\subsection{Suspiciously Structured Entropy}   \label{HSE}

Based on the reasoning above, previous research has used high mean entropy as an indicator of encryption or compression.  However, malicious contents, when concealed in a sophisticated manner,  may not be detectable through simple entropy statistics, such as mean file entropy.   Malware writers sometimes try to conceal hidden encrypted or compressed code that they introduce in creating files such as parasitic malware; for instance, they may add additional padding (zero entropy chunks), so that the file passes through high entropy filters. However, files with concealed encrypted or compressed segments tend to vacillate markedly between native code, encrypted and compressed segments, and padding, with each segment having distinct and characteristic expected entropy levels.    Thus, the field of cybersecurity  has started to pay attention to files with \emph{highly structured entropy}~\cite{S}, \cite{baysa}, that is, files whose code flips between various distinguishing levels of entropy through the file.   

In order to automatically identify the degree of entropic structure within a piece of software, we represent each portable executable file as an \quotes{entropy stream.}  
The entropy stream describes the amount of entropy over a small snippet of code in a certain location of the file. 
The \quotes{amount} of entropic structure can then be quantified, such that we can differentiate, for example, between a low-structured signal with a single local mean and variation around that mean, versus a highly-structured signal whose local mean changes many times over the course of the file.

In this paper\footnote{This paper is a development of earlier research originally published in conference proceedings~\cite{wojnowicz}.  For a more comprehensive viewpoint, see~\cite{wojnowicz2}.}, we define \emph{suspiciously structured entropy} as a \emph{particular pattern of entropic structure} which matches those of malicious files.  To quantify the suspiciousness of the structured entropy within a piece of software, we develop the notion of a \quotes{Suspiciously Structured Entropic Change Score} (SSECS).   We first describe how to calculate SSECS as a single predictive feature, and analyze its performance in malware detection.  We then generalize this feature to large-scale malware detection tasks. The derivation of the SSECS feature depends upon the notion of a wavelet transform, which we now briefly review. 

\subsection{Brief Overview Of Wavelets} \label{BOOW}

The Wavelet Transform is the primary mathematical operator underlying our quantification of structurally suspicious entropy.  The Wavelet Transform extracts the amount of \quotes{detail} exhibited within a signal at various locations over various levels of resolution \cite{Nason}.  In essence, it transforms a one-dimensional function of \quotes{location} (in our case, file location) into a two-dimensional function of \quotes{location} and \quotes{scale.}
By using the output of the wavelet transform (the so-called \quotes{wavelet coefficients}), it is possible to obtain a series of coarse-to-fine approximations of an original function. These successive approximations allow us to determine the multi-scale structure of the entropy signal, in particular the \quotes{energy} available at different levels of resolution. 
 
For this paper, we apply Haar Wavelets, which is a particularly simple family of wavelets whose members are piecewise constant.   The Haar Wavelet Transform projects the original entropy signal onto a collection of piecewise constant functions which oscillates as a square wave over bounded support (i.e., the functions assume non-zero values only on certain bounded intervals). Since these piecewise constant functions have supports which vary in their scale (width) and location, the resulting projections describe the \quotes{detail} within the signal at various locations and resolutions. 

More specifically, the Haar Wavelet Transform is based upon the so called \quotes{mother function}, $\psi(t)$, defined by:
\[ \psi(t) = \begin{cases}  1, & t \in [0,1/2) \\ -1, & t \in [1/2,1) \\ 0, & \text{otherwise} \end{cases} \]
a very simple step function.  Given the Haar mother function $\psi(t)$, a collection of dyadically scaled and translated wavelet functions $\psi_{j,k}(t)$ are formed by:
\begin{equation}
\label{eq_wavelet_functions}
\psi_{j,k}(t) = 2^{j/2} \psi (2^j t-k) 
\end{equation}
where the integers $j,k$ are scaling parameters.  The dilation parameter $j$ indexes the level of detail or resolution, and the translation parameter $k$ selects a certain location within the signal to be analyzed.  Note that as the scaling parameter $j$ increases, the function $\psi_{j,k}$ applies to (is non-zero over) successively finer intervals of the signal.   Some example Haar wavelet functions are shown in Figure 1.

\begin{figure}
\centering{ \includegraphics[width=.8\textwidth]{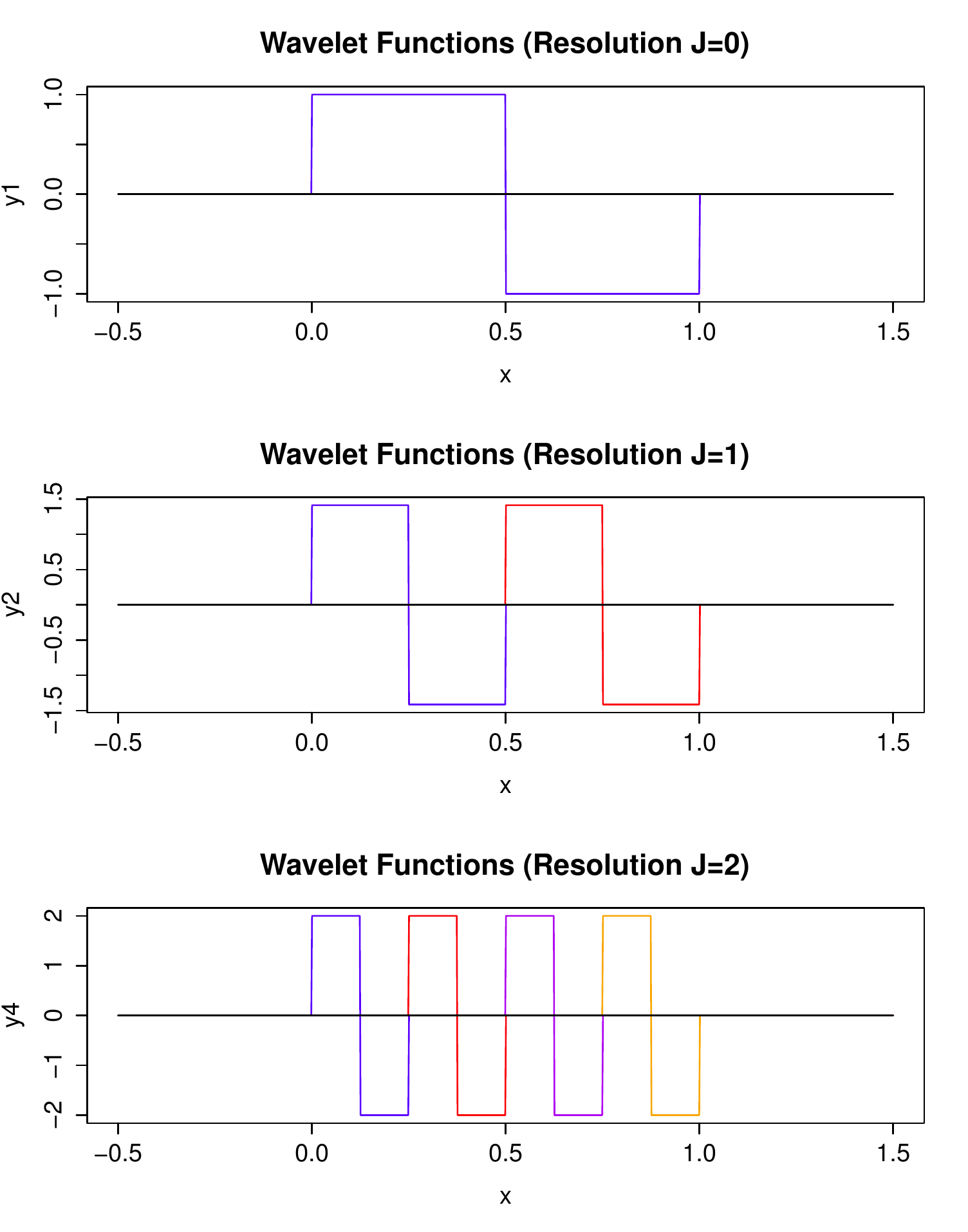}
\label{wavelet_functions}
\caption{ \emph{Examples of Haar wavelet functions.} Here we show some Haar wavelet functions over the unit interval.  Each colored square wave represents (the non-zero part of) a different wavelet function. The Haar wavelet functions are defined in Equation~\ref{eq_wavelet_functions}.  In particular, we plot wavelet functions for resolution levels $j=0,1,2$ and locations $k=0,..,j$. These wavelet functions are used as filters to pick up the magnitude of entropic change in a piece of software at different levels of resolution and in different file locations.}
} \end{figure}

Given a signal $x(t)$ where $t=1,\hdots,T$, we first rescale the signal so that the first observation occurs at time $t=0$ and the final observation occurs at time $t=1$.  Then, the so-called \quotes{mother wavelet coefficient} at scale $j$ and location $k$ is given by the inner product of the signal with the wavelet.  Since we are dealing with discrete signals, the inner product takes the form:
\[d_{j,k} = <x, \psi_{j,k}> = \ds\sum_{t=1}^T x(t) \psi_{j,k} (t),\]
 One interpretation of this coefficient is that it gives the (scaled) difference between local averages of the s
ignal across neighboring chunks or bins.  The size of the neighboring chunks is determined by the scaling parameter $j$. 
 
The family of mother wavelet coefficients, $\{d_{j,k}\}$, enable a \quotes{Multi-Resolution Analysis} (MRA) of the signal $x(t)$. In particular, the signal $x(t)$ can be decomposed into a series of approximations $x_j(t)$ , whereby each successive approximation $x_{j+1}(t)$ is a more detailed refinement of the previous approximation, $x_j(t)$.  The functional approximations are obtained through the wavelet coefficients by the formula:
\begin{equation}
\label{functional_approximations}
x_{j+1}(t) = x_j(t) + \ds\sum_{k=0}^{2^j-1} d_{j,k} \psi_{j,k}(t)  
\end{equation}
where  $x_0(t)$, the coarsest-level functional approximation,  is the mean of the full signal.  Thus, the collection of mother wavelet coefficients $\{d_{j,k}\}$ store the \quotes{details} that allow one to move from a coarser approximation to a finer approximation.  Examples of successive functional approximations, in the context of software entropy signals, are shown in Figure 2. 

\begin{figure}
\centering{ \includegraphics[width=.9\textwidth]{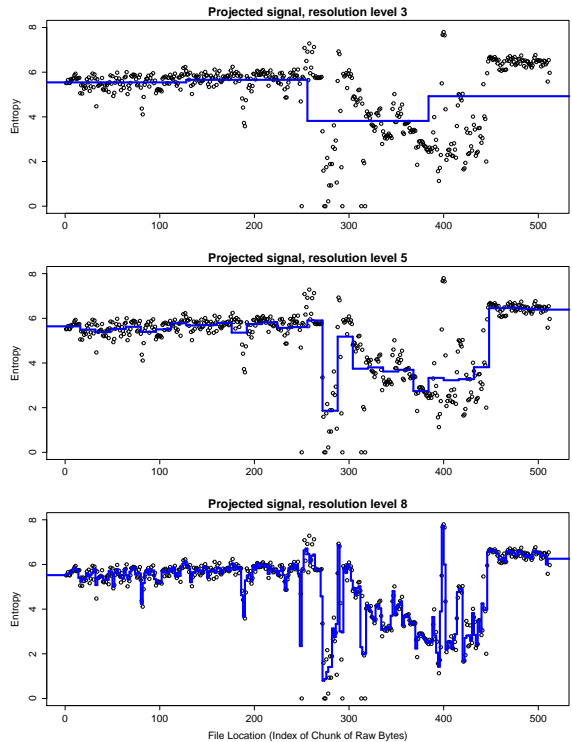}
\label{projected}
\caption{ \emph{Wavelet-based functional approximations to a software's entropy signal at different levels of resolution.}  Here, we show the entropy signal from a single Portable Executable (PE) file projected onto Haar father wavelet space at different levels of resolution ($j \in \{2,5,8\}$ from Equation~\ref{functional_approximations}).  In general, each successive functional approximation adds the incremental detail provided at that level of spatial resolution, compared to the next-most-coarse level of spatial resolution, and does so across various spatial locations.}
} 
\end{figure}

Using the wavelet transform, it is possible to \quotes{summarize} the overall amount of detail in a signal at various levels of resolution.  The total amount of detail at a particular ($j$th) level of resolution is known as the \emph{energy} at that level of resolution:
\begin{equation}
\label{Energy}
E_j = \displaystyle\sum_{k=1}^{2^{j-1}} ( d_{jk})^2  
\end{equation}

The distribution of energy across various levels of resolution is known as an \emph{energy spectrum}.     Note that the energy at resolution level $j$ is just the squared Euclidean norm of the vector of mother wavelet coefficients from resolution level $j$.  After this step, we have reduced the original signal of size $T=2^J$ (and resultant wavelet vector of size $T-1$) to a vector of $J$ elements, where each element represents the amount of \quotes{energy} at a single level of resolution.

\subsection{Wavelet-Based Classifiers}

 The energy spectra of signals have been very useful features for classifiers such as neural networks. In fact, this combined strategy, whereby the coefficients from a discrete wavelet transform are used as node activations in a neural network, is referred to as a wavelet neural network (\emph{WNN}) strategy (see e.g. \cite{Pati},~\cite{subasi}).   Using WNN's, researchers have been able to automatically classify lung sounds into categories (crackles, wheezes, striders, squawks, etc.) \cite{Lung}, to automatically determine whether brain EEG scans originated from healthy patients, patients with epilepsy, or patients who were in the middle of having a seizure \cite{Omerhodzic}, or to automatically determine whether EMG signals collected from the bicep originated from patients who were healthy, suffering from myopathy, or suffering from neurogenic disease \cite{EMG}.  

We refer to the overall strategy of using wavelet coefficients as features in a classifier as a \emph{Wavelet-Based Classifier} strategy.  We prefer this term over \emph{WNN}, which, although well-established in the literature, is specific to neural network classifiers.  
Indeed, in this paper, we choose logistic regression (both standard and regularized) rather than a neural network to model our data, because the logistic regression model provides an atomic analysis of the relationship between the wavelet-based features and classification categories.    



\subsection{Suspiciously Structured Entropic Change Score (SSECS)} \label{SSES}

The initial fundamental problem with applying wavelet-based classifiers to malware analysis is that executable files out \quotes{in the wild} have different lengths.   This contrasts with controlled observational situations, e.g. those described above, which produce signal samples of fixed length that are held constant across the data set.  In controlled observational situations, all samples will produce the same number of features, J, and variation across these set of J features can be immediately associated with a classification variable in a straightforward manner, for example by setting the input layer of the neural network to have J activation notes.  

However, in uncontrolled observational contexts, signal lengths can differ wildly from sample to sample.  
Imagine, for instance, comparing signal A of length 32 (so J=5, and if $E_{f,j}$ represents the energy at resolution level $j=1,\hdots,J$ for portable executable file $f$, we would have $E_{a,1}, \hdots, E_{a,5})$ with signal B of length 256 (so J=8, and we have $E_{b,1}, \hdots, E_{b,8}$).    How should we compare these two files?  



Our solution to this problem, for smaller data sets\footnote{A second solution, for larger datasets, is described in Experiment 2.}, is to transform each file's $J$-dimensional energy spectrum into a single scalar feature, a $1$-dimensional \quotes{Suspiciously Structured Entropic Change Score} (SSECS).   The computation of SSECS is a two-step process: first, we compute the wavelet-based energy spectrum of a file's entropy signal, and second, we compute the file's malware propensity score from that energy spectrum.    
In our case, we fit a logistic regression model to the binary classification response (malware or not) which uses these wavelet energy features as predictor variables. We fit $J$ separate regression models, one for each file size grouping.   Given the Energy Spectrum $\{E_{f,j}\}$, which is the set of wavelet energies for each resolution level $j=1,\hdots,J$ of portable executable file $f$, the logistic regression model estimates $\widehat{P}_f$, the predicted probability that file $f$ is malware, by the formula 
\[ \widehat{P}_f = \df{1}{1+exp[ -\beta_0 + E_{f,j} \cdot \beta^{(J)} ]} \]
where $\beta_j^{(J)}$ is a model parameter, known as a \quotes{logistic regression coefficient}, from the $J$th  logistic regression model.  This number, $\widehat{P}_f$ is what we refer to as the SSECS. 



\section{Experiment 1: Analyzing and evaluating the predictive performance of a 
 single wavelet-based feature}

In Experiment 1, we attempt to assess the predictive value of SSECS as a single feature describing potentially suspicious variation in software entropy.   In particular, as discussed in Section~\ref{expt1_method}, the wavelet-based feature is constructed in an attempt to describe the ``suspiciousness" of a piece of software's entropy signal when that entropy signal is re-represented, through a wavelet transform, in terms of entropic change distributed across different levels of spatial resolution.  

\subsection{Data}

Data are a set of n=39,968 portable executable files from a Cylance repository.  19,988 (50.01\%) of these files were known to be malicious, and the remaining files were benign.  These files were collected ``from the wild," and thus highly heterogenous.   For example, the ``malware" category contains different types of malicious software (e.g. viruses, Trojan horses, spyware, backdoors, bots, and ransomware -- but not adware.)

\subsection{Method} \label{expt1_method}

\subsubsection{Constructing  the entropy stream}
To compute the entropy of an executable file, the original file, represented in hexadecimal (00h-FFh), is split into non-overlapping chunks of fixed length, typically 256 bytes.  For each chunk of code, the entropy is then computed using the formula below:
\begin{equation}
\label{H}
H(c) = - \displaystyle\sum_{i=1}^m p_i(c) \log_2 p_i(c), 
\end{equation}
where $c$ represents a particular chunk of code, $m$ represents the number of possible characters (here, n=256), and $p_i$ is the probability (observed frequency) of each character in the given chunk of code.  The entropy for any given chunk then ranges from a minimum of 0 to a maximum of 8.  

\subsubsection{Computing the Suspiciously Structured Entropic Change Score (SSECS)} \label{SSES_Compute}


The procedure for computing the suspiciously structured entropic change score (SSECS) is as follows:

\begin{enumerate}[leftmargin=0cm,itemindent=.5cm,labelwidth=\itemindent,labelsep=.2cm,align=left]

\item[1)] \underline{Partition data set by size}:  Group sampled files into $j=\{1,\hdots,J\}$ groups, where $j={\lfloor log_2 T \rfloor}$ and T is the length of the file's entropy stream:

\item[2)] \underline{Iterate}:  For all files which fall into the  $j$th length group
\begin{enumerate}

\item[2a)]  \underline{Compute Haar Discrete Wavelet Coefficients}:  The discrete wavelet transform takes as input a discrete series of size $T=2^J$ observations. Because the transform requires the series to have a dyadic length, if the number of observations in the executable file's entropy stream is not an integer power of 2, we right-truncate the series at value $2^{\lfloor log_2 T \rfloor}$. The so called \quotes{mother} wavelet coefficients,  $d_{jk}$, describe the \quotes{detail} at successively fine-grained resolutions.   
In particular, the mother wavelet coefficients are indexed such that $j \in \{ 1, \hdots, J\}$ represents the resolution level, ordered from coarse-grained to fine-grained, and $k \in \{ 1, \hdots, K=2^{j-1}\}$ represents the particular location (or bin) of the entropy signal at that resolution level.  At each resolution level $j$, the signal is divided into $N_j =2^{j-1}$ non-overlapping, adjacent bins such that each bin includes $B_j = 2^{J-j}$ observations. Note that the number of bins, K, increases as $j$ increases to finer resolutions.   The mother wavelet coefficient at index $(k,j)$ is then given by: 

\begin{equation}
d_{kj}  = \df{1}{s_j} \bigg( \ds\sum_{i=(2k-1)Bj+1}^{2kBj} y_i  - \ds\sum_{i=(2k-2)Bj+1}^{(2k-1)Bj}  y_i \bigg) 
\label{dkj}
\end{equation}

where the scaling factor is $s_j =(\sqrt{2})^{J-j+1} $  and is necessary for the wavelet transform to preserve the size (norm) of the signal. There are T-1 mother wavelet coefficients.  

\item[2b)] \label{CWES} \underline{Compute Wavelet Energy Spectrum}: The \emph{wavelet energy spectrum} summarizes the \quotes{detail} or \quotes{variation} available at various resolution levels.  The energy spectrum is computed as a function of the mother wavelet coefficients, $d_{jk}$.  In particular, the \quotes{energy}, $E_j$, of the entropy stream at the $j$th resolution level is defined by Equation~\ref{Energy}.   Given a particular executable file's entropy stream, we refer to its distribution of energy over different resolutions the file's \quotes{energy spectrum.}

\item[2c)]  \underline{Compute Wavelet Energy Suspiciousness}:  Now we use the wavelet energy spectrum to  determine the \quotes{propensity} of each file to be malware (i.e., its suspiciousness).  Computing this propensity requires training.  We use 5-fold validation.

\begin{enumerate}
\item[2c1)] \underline{Partition The Current Sample Of Files:} Split the entire set of $F_J$ files which are of the appropriate size into 5 mutually exclusive subsets $F_J^1, \hdots, F_J^5$, each of which represents exactly 20\% of the entire sample.     
\item[2c2)]  \underline{Iterate:}  For each subset $F_J^i$, where $i \in \{1,\hdots,5\}$
\begin{enumerate}
  \item[2c2a)]  \underline{Fit a logistic regression }:  Fit a logistic regression model on the other four subsets $\{F_J^k : k \neq i  \}$, where the model fits the class variable (malware or not) as a function of the wavelet energy spectrum.  The logistic regression model will produce a set of beta coefficients to weigh the strength of each resolution energy on the file's probability of being malware.
  \item[2c2b)]  \underline{Calculate malware propensity}: Use the logistic regression model above to then make a prediction about files in subset $F_J^i$. In particular, use the model learned in step 1c2a to calculate the predicted probability that each file in set $F_J^i$ is malware, given its wavelet energy spectrum.   This malware propensity (i.e., predicted malware probability) lies within the interval $[0,1]$, and is what we call the Suspiciously Structured Entropic Change Score (SSECS). 
\end{enumerate}
\end{enumerate}
\end{enumerate}
\end{enumerate}


\subsection{Results}

\subsubsection{Suspicious Patterns of Entropic Change in A Single File Size Group} \label{SLM}

How does the model transform these wavelet energy spectra into predictions about whether the file is malware (that is, into a Suspiciously Structured Entropic Change Score)? To illustrate, we consider the subset of n=1,599 files in our corpus belonging to file size group $J=5$.  Because these files can be analyzed at $J=5$ different spatial resolutions, we extract 5 features from each file, with each feature representing the energy at one level of spatial resolution in the file's entropy stream. 

For illustrative purposes, we begin by analyzing the wavelet energy spectrum for two files from this size category, as they embody more general trends in the energy patterns of malicious versus clean files.    Figure 3 shows wavelet-based functional approximations for two different entropy streams. The left column of the plot depicts the entropy signal from File A, which is legitimate software, whereas the right column of the plot depicts the entropy signal from File B, which is malware.   Reading these columns from top to bottom, we see that the wavelet transform produces successively detailed functional approximations to these files' entropy signals.   The title above each subplot shows the wavelet energy, as computed in Equation~\eqref{Energy} in the text, of the signal at a particular spatial resolution level.  The wavelet energy is simply the sum of the squares of the scaled differences in the mean entropy levels, where the differences are only taken between even/odd index pairings (i.e. the algorithm takes the differences $mean_{bin2}-mean_{bin1}, mean_{bin4}-mean_{bin3}$, and so forth).    Thus, we can gain some visual intuition about how the energy spectra can be derived from these successive functional approximations.    

Based on this entropic energy spectrum decomposition (or distribution of energy across various levels of spatial resolution), the model believes that File A is legitimate software, whereas File B is malware.  Investigating this conclusion, we see that these two files have radically different wavelet energy distributions across the 5 levels of spatial resolution.  The legitimate software (File A) has its \quotes{entropic energy} mostly concentrated at finer levels of resolution, whereas the piece of malware (File B) has its \quotes{entropic energy} mostly concentrated at coarser levels of resolution.    For the clean file, the energy in the entropy stream is concentrated at the resolution levels $j=4$ and $j=5$ (where the energy is 34.5 and 23.84 squared bits, respectively).  For the dirty file, the energy in the entropy signal is concentrated at coarser levels of analysis, peaking especially strongly at level $j=2$ (where the energy is 139.99 squared bits).  

\begin{figure*}
\centering{ \includegraphics[width=.82\textwidth]{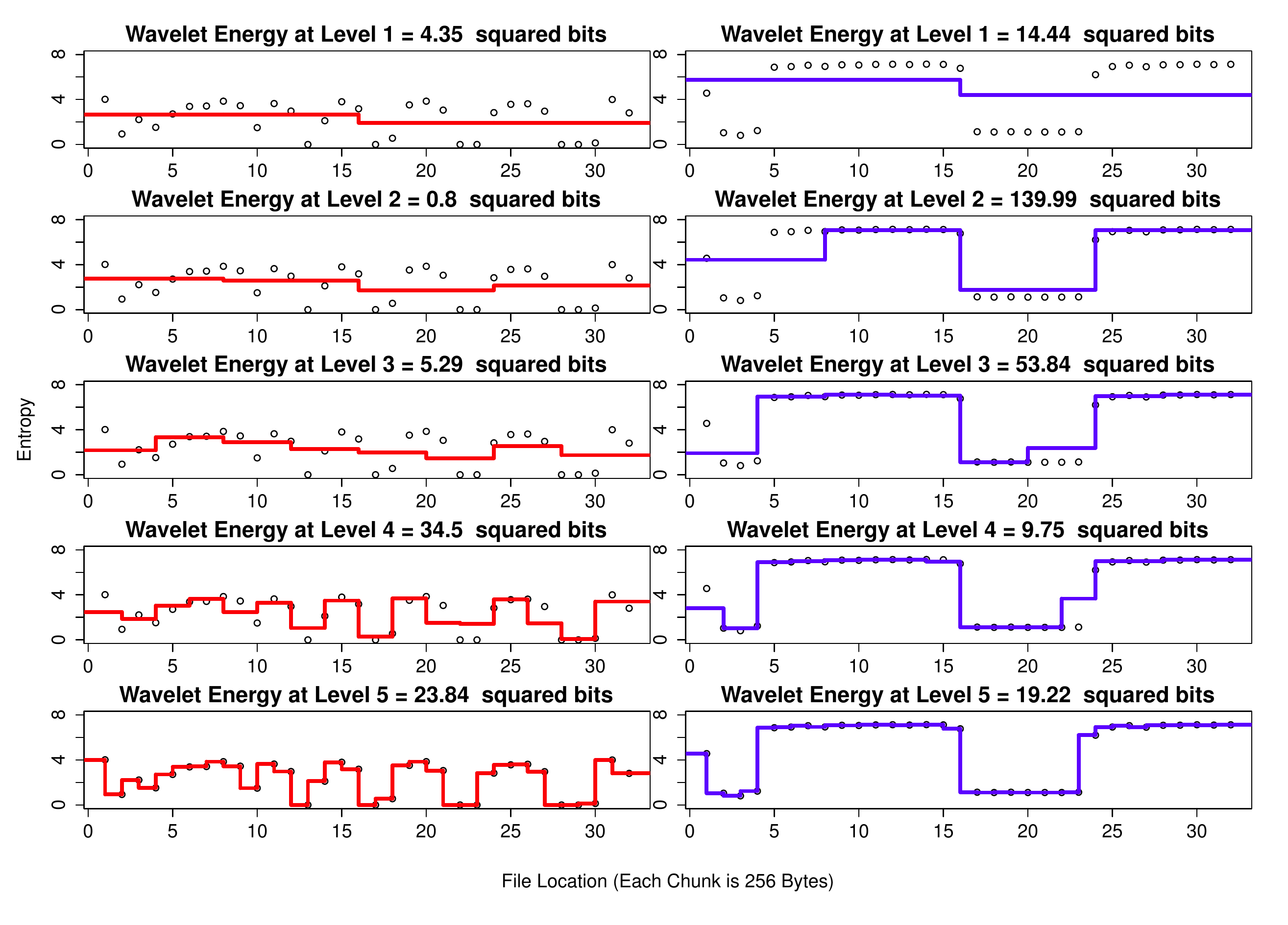}
\label{EnergyPlot}
\caption{Wavelet-based functional approximations, and the corresponding wavelet energy spectrum, for the entropy signals of two representative portable executable files from one file size group.}
} \end{figure*}

The fit of the logistic regression model (for both raw and normalized features) is summarized in Table 1.   Note that for the entire table, numbers outside the parentheses represent results for the normalized features, whereas numbers inside the parentheses represent results for raw features.   The two \quotes{Energy} columns list the energy at all five levels of spatial resolution for these two files.    The \quotes{Value of $\beta_j$} column describes the estimated beta weight in a logistic regression fitting file maliciousness to the five wavelet energy values, based on a corpus of n=1,599 files. The \quotes{P-value} column describes the probability of getting the test statistic we observed (not shown, it is a function of the data) under the hypothesis that there is no relationship between energy at that level and file maliciousness.    The codes are: $*=p<.05, **=p<.01, ***=p<.001, ****=p<.0001, ******=p<.00001$. The \quotes{Malware Sensitivity} represents the estimated change in the odds that a file is malware associated with an increase of one unit in the corresponding feature.  It is calculated by $(e^{\beta}-1) \times 100\%$.  For the normalized values (those outside the parenthesis), an increase of one unit refers to an increase of one standard deviation.  

Based on these logistic regression beta weight ($\beta_j$) values, we see that the two sample files from Figure 3 are indeed representative of a larger trend: having high energy at spatial resolution levels 1,2 and 3 (the coarser levels) is associated with a higher probability of the file being malware (since those $\beta_j$'s are positive), whereas having high energy at levels 4 and 5 (the finer levels) is associated with a lower probability of the file being malicious (since those $\beta_j$'s are negative).  Moreover, these associations appears to be reflective of trends in the larger population of files, since the p-values are largely strongly statistically significant.  This finding makes sense if artificial encryption and compression tactics tend to elevate moderate to large sized chunks of malicious files into high entropy states.


\begin{table*}\centering
\ra{1.3}
\footnotesize
\tabcolsep=0.05cm
 \begin{adjustbox}{max width=\textwidth}
\begin{tabular}{@{}rrr c rr c rrr@{}}\toprule
 \multicolumn{3}{c}{$Resolution$} & \phantom{ab}& \multicolumn{2}{c}{$Energy \; Spectra$} &  \phantom{ab}& \multicolumn{3}{c}{$Statistical \; Model \; For \; File \; Size \; J=5 $} \\
 $Level$ & $\# \; Bins$ & $Bin \; Size$ && $File \; A$ & $File \; B$ & & $Value \; of \; \beta_j$ &$P-value$ & $Malware \; Sensitivity$ \\ \midrule
  $1$ & 2 & 16 && -0.39 (4.35) & -0.01 (14.44)& & 0.448 (0.017) & {\bf *****} & +56.5\% (+1.7\%) \\
  $2$& 4 & 8 && -0.79 (0.80) & 6.27 (139.99) & &0.174 (0.008) & {\bf *}& +19.0\% (+0.89\%)\\
 $3$ & 8 & 4 && -0.48 (5.29) & 2.18 (53.83)  & &0.847 (0.046) & {\bf *****} & +133.2\% (+4.74\%) \\
 $4$ & 16 & 2 && 1.42 (34.50) & -0.37 (9.75) & &-0.106 (-0.008) & n.s. & -10.0\% (-0.75\%) \\
 $5$ & 32 & 1 && 1.77 (23.84) & 1.19 (19.22) & &-0.240 (-0.030) & {\bf **} & -21.4\% (-2.99\%) \\  \bottomrule
\end{tabular}
\end{adjustbox}
\caption{Investigating the relationship between the entropic wavelet energy spectrum and maliciousness for files in one size group.  }
\label{SLT} 
\end{table*}


\subsubsection{Suspicious Patterns of Entropic Change Across  All File Size Groups}

 
Do the trends found in the single level analysis of $n=1,599$ files hold up in the full corpus of $n=39,968$ files? In particular, regardless of file size, can we corroborate the simply stated conclusion that  \quotes{malware tends to concentrate entropic energy at relatively coarse levels of spatial resolution?}  And if so, where is the dividing line between  \quotes{coarse} and \quotes{fine}?  

In Figure 4, we summarize the results of logistic regression models fits across all file size groupings.
   The plot shows logistic regression beta coefficients for determining the probability that a portable executable file is malware based upon the magnitude of file's entropic energy at various levels of spatial resolution within the code.       Positive betas (red colors) mean that higher \quotes{entropic energy} at that resolution level is associated with a greater probability of being malware.  Negative betas (blue colors) mean that higher  \quotes{entropic energy} at that resolution level is associated with a lower probability of being malware.  For both colors, stronger intensities represent stronger magnitudes of the relationship between entropic energy and malware.   Mathematically, the dot product between a file's energy spectrum and these beta weights determine the fitted probability that the file is malicious.    Thus, the  \emph{Danger Map} interpretation can be interpreted as follows:  For any file size grouping (or row), files that have high energies in the red spots and low energies in the blue spots are significantly more likely to be  \quotes{dangerous.}  Conversely, files that have low energies in the red spots and high energies in the red spots are significantly more likely to be  \quotes{safe.}  

Taking this \emph{Danger Map} into consideration, we draw the following conclusions:
\begin{itemize}
\item To a first approximation, the full analysis supports the  \quotes{coarse-energy-is-bad, fine-energy-is-good} mantra (observed in Section~\ref{SLM}'s analysis of a single file-size group).    Visually, most diagonal elements of the matrix are blue (and also more blue than the off-diagonals).   Thus, across most file sizes, high energies at the finest-level of spatial resolution appear to be indicative of file legitimacy, and high energies at coarse levels of spatial resolution are often associated with suspiciousness.  
\item However, what qualifies as a suspicious pattern in the wavelet decomposition of a file's entropy stream appears to be more
 complex than the simplistic summary above.  
 For example, the appearance of the double diagonal bands in blue suggest somewhat regular vacillations in terms of how  \quotes{suspicious} high entropic energy would look at various levels of spatial resolution.  We find that the particular patterning depicted in the Danger Map provides a statistically significantly better  description of malware than random (baseline-informed) guessing alone. Likelihood ratio tests comparing the fit of the size-specific models (where the beta coefficients of each size-specific model are given by the specific colorings in the corresponding row of the Danger Map) versus the fit of models with no features (interpretable as a uniform color across rows, where the intensity of the color is determined by baseline malware rates, independent of the wavelet energy spectrum) yield the test statistics below.  Moving from bottom (J=3) to top (J=15) of the figure, we have:

\scriptsize
\begin{align*}
& \Chi^2(3)=198.36, &&\Chi^2(4)= 563.51, &&\Chi^2(5)= 257.52, \\
& \Chi^2(6)= 235.09, && \Chi^2(7)=150.11, &&\Chi^2(8)= 585.57, \\
 &\Chi^2(9)= 662.22, && \Chi^2(10)= 283.24, &&\Chi^2(11)= 385.33, \\
 &\Chi^2(12)= 305.04, && \Chi^2(13)= 233.39, &&\Chi^2(14)=116.17, \\
 &\Chi^2(15)= 61.88, && \\
\end{align*}
\normalsize
All of these test statistics achieve statistical significance at the $\alpha=.05$ level.    Moreover, even after a conservative Bonferroni's correction for simultaneous hypothesis testing (of 10 null hypotheses), we can still reject the null hypothesis of a uniform color across rows for each spatial resolution except spatial resolution level 9.   This finding suggests that the distribution of colors in the ``Danger Map" of Figure 4, while not sufficiently simplistic to be easily verbalizable, is unlikely to be obtainable by random chance.\footnote{ We reject the null hypothesis that the colors in each row are uniform, and this rejection is consistent with the hypothesis that the complex patterns of colors are meaningful in predicting malware. However, we point out for the sake of completion that this finding is also consistent with simpler but more specific hypotheses, such as that the right-most off-diagonal cell is driving the result.  Ideally, a more sophisticated statistical model, well-tailored to the structure of a multi-resolution dataset, would be applied here 
to tease apart these remaining possibilities. }  
 \end{itemize}

\begin{figure}
\label{DangerMapPlot}
\centering{ \includegraphics[height=2.5in]{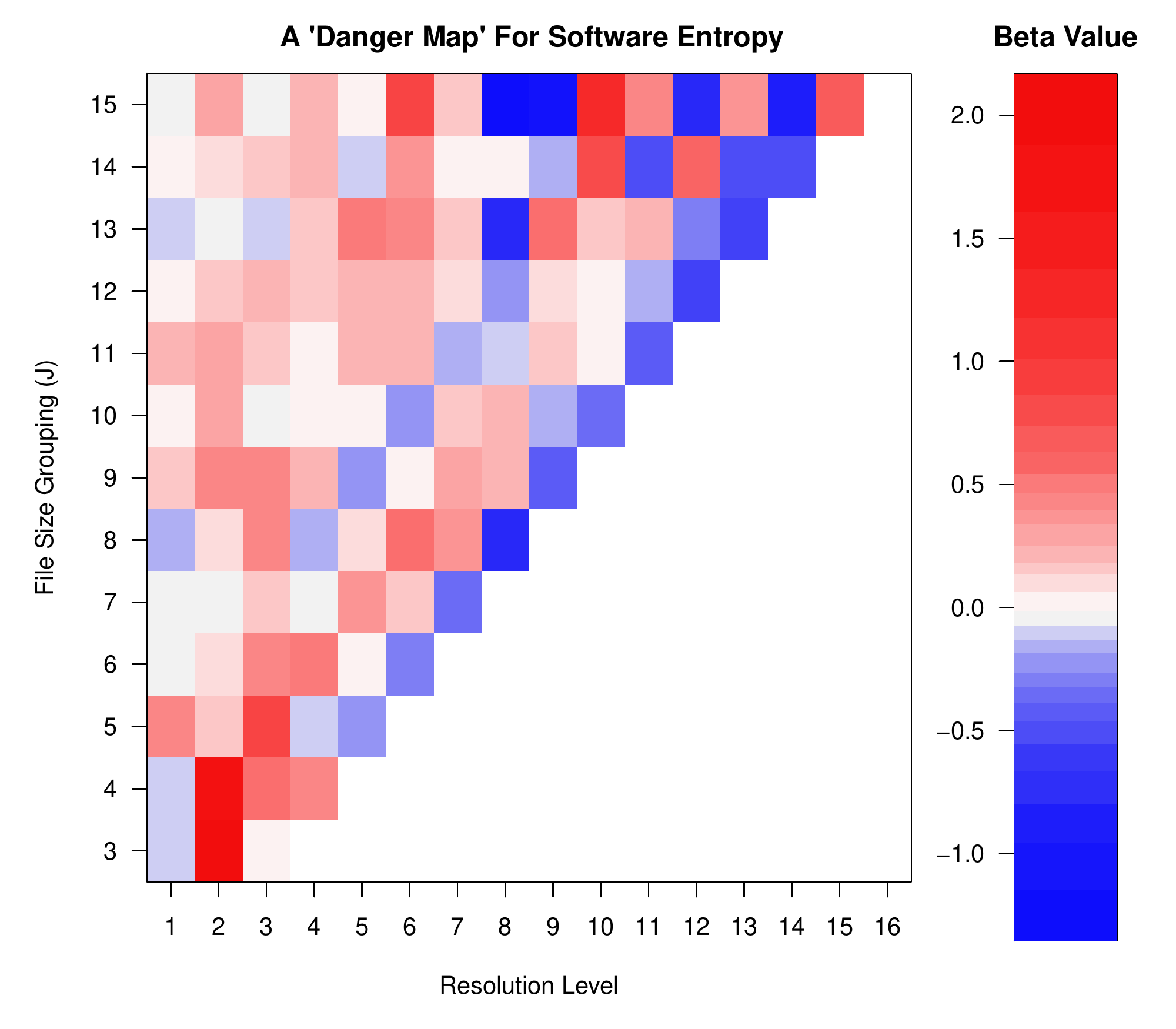}
\caption{ \emph{A  ``Danger Map" for entropy patterns within a piece of software.}  The danger map is derived from a statistical model of malware classification which learns suspicious patterns inherent within each software's entropy streams.  In particular, a wavelet decomposition of these entropy streams reveals the entropic energy at various levels of resolution.  The plot shows logistic regression beta coefficients for determining the probability that a portable executable file is malware based upon the magnitude of file's entropic energy at various levels of resolution within the code. } 
}
 \end{figure}

\subsection{Predictive performance of the single wavelet feature} \label{SSES_Results}

How can we use the information distributed across the \quotes{Danger Map} to construct a single number which could score a piece of software's suspiciousness based on the wavelet decomposition of its entropy signal?  
We studied the predictive performance of SSECS in identifying malware by constructing a hold-out test set of $n=7,991$ files and found:

\begin{enumerate}
\item  \emph{SSECS as a single feature improved predictions of malware}, within a balanced sample of malware and legitimate software, from 50\% to 68.7\% accuracy.   This makes SSECS a particularly impressive feature, considering that most machine learning models of malware consist of millions of features.

\item  \emph{SSECS provides predictive information beyond what is contained in a mean entropy feature}.    A model with mean entropy as a single feature achieved 66.2\% predictive accuracy.   Thus, mean entropy is indeed also an impressive single predictor of malware (perhaps not surprisingly given its prevalence in the literature).    However, unlike mean entropy, the wavelet energy spectrum detects suspicious patterns of entropic \emph{change} across the code of the executable file.   We found that a 2-feature model which includes both mean entropy and SSECS achieves 73.3\% predictive accuracy (so adding wavelet-based information to the model yields a 7.1\% boost in predictive accuracy beyond what is obtained by mean entropy alone).   

\item  \emph{SSECS provides predictive information beyond what is contained in a ``standard deviation of entropy" feature}.     A skeptic might ask: why not simply use standard deviation, a more commonly used and more computationally straightforward measure of variation?  Standard deviation is useful, but a relatively cruder measure of variation, as it operates on only a single spatial scale.  Indeed, a 2-feature model which includes both mean entropy and standard deviation achieves merely 70.4\% predictive accuracy.  

\end{enumerate}

\section{Experiment 2: Larger-scale detection of parasitic malware}

In Experiment 1, we evaluated the predictive value of a single wavelet-based feature that describes how software's entropic shifts are distributed across multiple spatial scales.   We found that this feature can exploit valuable information from a software's entropy signal which is relevant to malware status and which goes beyond the predictive value of the most commonly used entropy measures, mean entropy, as well as a potentially conceptually simpler measure of entropy variation,  entropy standard deviation.  In Experiment 2, we apply a broader system of wavelet-based features to a larger-scale malware prediction task.    In particular, the task is to identify parasitic malware from a large corpus of otherwise good files.  Parasitic malware generally infects existing files on a user's system, and the infected part of the file typically conceals itself through encryption or compression.  Thus, if wavelet decomposition of software entropy indeed yields features which successfully track the presence of suspicious chunks of encrypted or compressed code, then these features should be particularly valuable for a parasitic detection task.  

\subsection{Data}
Data were 699,121 samples of Portable Executable (PE) files from a Cylance repository.  Of these samples, 17,605 files (2.51\%) were parasitic malware, and the remaining files were legitimate software.  We randomly selected 80\% of the dataset for training, and the remaining 20\% were allocated to the test set. 

\subsection{Method} \label{expt2_method}
To validate the utility of wavelet features in distinguishing parasitic malware from clean software,  we compared four models (in the sense of types of features extracted from executable files to feed into a machine learning classifier):
\begin{enumerate}
\item \emph{Strings Model}: A strings-only model is a common way to build features for a machine learning classifier~\cite{Schultz}.  Thus, we extract the $P_1=1,117,127$ most common strings observed in our corpus and use them as binary features in a predictive model. 
\item \emph{Strings+Wavelet Model}: We would like to investigate if wavelet-based features can add predictive value to a strings only model.  Because of the relatively large-scale size of the dataset ($\approx 20\times$ the size of Experiment 1), we streamline the feature generation process.  Rather than computing SSECS, the energy spectrum suspiciousness score, which requires a nested modeling step, we follow the feature generation algorithm of Section~\ref{expt1_method} only up to Step~\ref{CWES}, computing the wavelet energy spectrum.   We then represent the wavelet energy spectrum separately for each file size group. In particular, a sample with $T$ points in its entropy stream will have $J=\lfloor \log_2{T} \rfloor$ features in its wavelet energy spectrum.  If $J_{max}$ is the maximum observed value of $J$ in the dataset, then there are $\sum_{J=1}^{J_\text{max}} J = \df{J_{max}(J_{max}+1)}{2}$ features, where any given sample with $T$ points in its entropy stream will only have non-zero values for $J=\lfloor \log_2{T} \rfloor$ of these features (namely, for the part of the vector that corresponds to its filesize group).   Although obviously this procedure creates a huge proliferation of features relative to the single SSECS feature studied in Experiment 1, the procedure is more informative and becomes more feasible as more data is collected, while simultaneously streamlining the modeling pipeline for larger datasets.  Finally, we bin the wavelet energy spectrum features, which are originally continuous, to create a sparse binary dataset.  In this way, we obtain 24,009 binary features derived from the wavelet energy spectrum.  After adding in the strings as well, the Strings+Wavelet model includes $P_2=1,141,136$ binary features. 
\item \emph{Strings+Entropy+Wavelet Model}:  The wavelet features capture some information about the entropy signal, but it is incomplete.  For example, the wavelet energy spectrum describes \emph{variation} at multiple levels of resolution, but ignores \emph{first-order} information (i.e., measures of central tendency, such as the mean).   Thus, in an attempt to construct a more powerful predictive model from strings and the entropy signal, here we add simple summary statistics about the entropy signal:  mean, standard deviation, signal-to-noise ratio, maximum entropy, percentage of the signal with \quotes{high} entropy ($\geq$ 6.5 bits), percentage of the signal with zero entropy, and length and squared length of the signal.    As these supplementary entropy features are relatively simple to compute, we obtain these measurements separately for each PE section.  As these features are also continuous, they are then binned through an internal binning process to create a sparse binary dataset.  This procedure creates 108,835 additional features to add to the strings model (24,009 derived from the wavelet energy spectrum, and 84,826 other entropy features). All together, this model contains $P_3=1,225,962$ binary features.
\item \emph{Strings+Entropy Model}:  In order to provide a more rigorous test of the value of the wavelet features, we create a fourth model which includes strings and the summary entropy features described above, but no wavelet features.    Our reasoning is that, even if the wavelet features improve the strings-only model, this improvement could, in theory, have been merely driven by the inclusion of some entropy information (or even file length).   By constructing this model, we can compare the performance of the Strings+Entropy+Wavelet model with the performance of the Strings+Entropy model to answer the question: do wavelet features provide additional predictive information that goes above and beyond the information inherent in summary entropy statistics (mean, max standard deviation, etc.)?   Thus, this model includes the 84,826 summary entropy features, but not the wavelet features.   All together, with the string features as well, this model contains $P_4=1,201,953$ features.
\end{enumerate}

Because we have a large number of predictors (up to $P_{max}=1,225,962$) relative to samples ($N=699,121$), we apply a ``logistic lasso" model (i.e. $\ell$1-penalized logistic regression) to perform classification and feature selection simultaneously.     Similarly to unregularized logistic regression, we can use the learned regression (or beta) weights as a proxy for feature importance. Since the features are all binary, each $\beta_j, \; j=1,\hdots,P$ can be interpreted as the increase in log odds that the file is malware which is associated with the $j$th feature ``turning on" (i.e. flipping from 0 to 1) and all other features staying constant.   Thus, features with large positive (respectively, negative) beta weights can be considered particularly strong predictors of goodness (respectively, badness).   In the results section, we explore properties of the most ``influential" features, defined as the collection of 100 features with the largest positive weights and 100 features with the largest negative weights.   As our purpose in this paper is to compare the effect of different feature subsets on predictive performance, and not to explore the predictive benefits of varying levels of sparsity in feature selection, we simply fix the sparsity parameter to 1.0.

\subsection{Results and Discussion}

\begin{figure*}
\label{ROCCurves}
\centering{ \includegraphics[width=.9\textwidth]{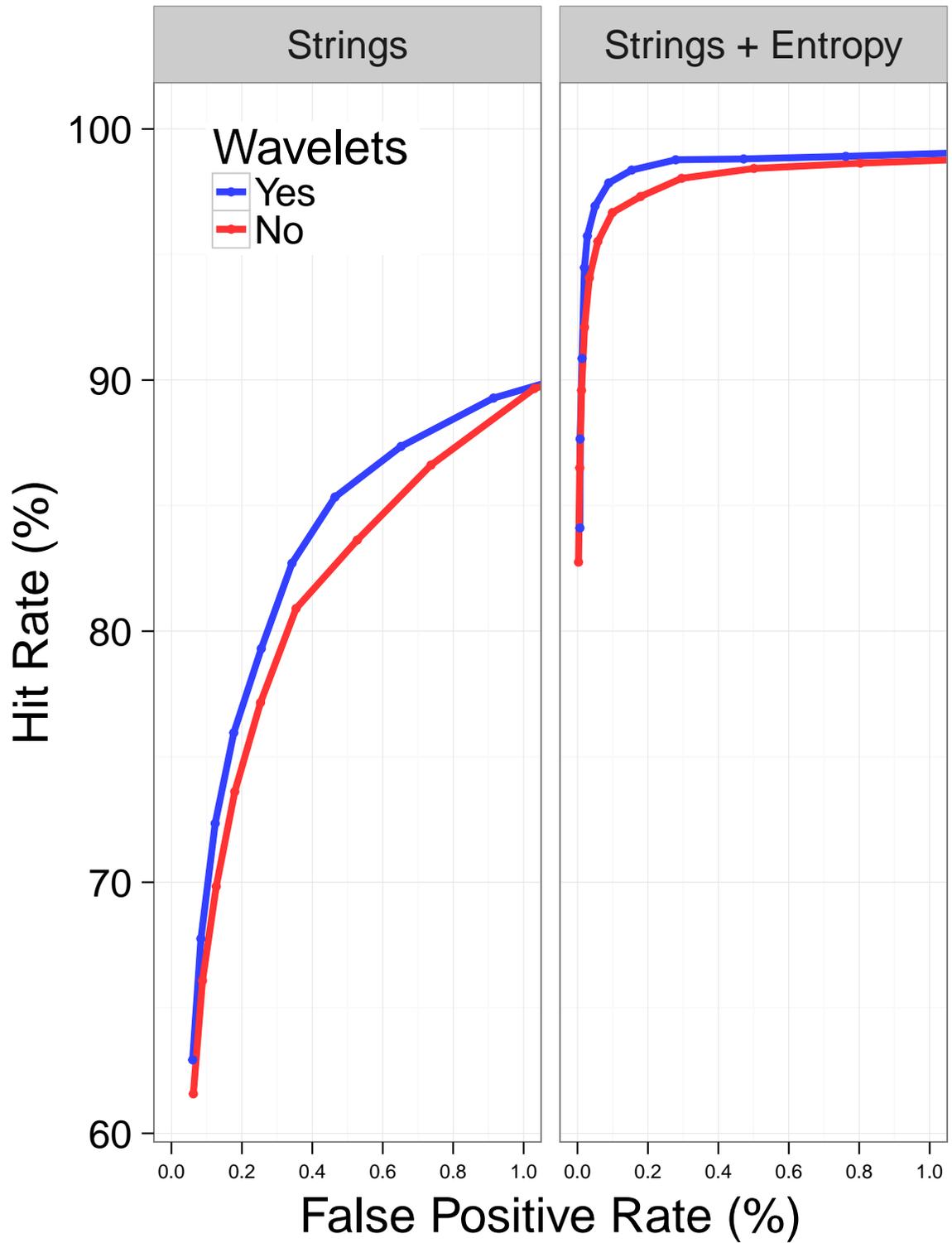}
\caption{ \emph{Performance boost on parasitic malware detection task caused by adding wavelet-based features to two different baseline feature processing methods.}  Performance here was measured as accuracy by a logistic lasso classifier on a hold-out test set of software samples. } 
}
 \end{figure*}

\begin{table*}\centering
\ra{1.3}
 \begin{adjustbox}{max width=\textwidth}
\begin{tabular}{@{}lll @{}}\toprule
 \phantom{abc} & \multicolumn{2}{c}{{\bf Predictive Accuracy (Test Set)}}  \\
{\bf   Model } & {\bf Parasitic Malware}& {\bf Clean Software }  \\ \midrule  
{\bf  Strings}   & 80.90\% & 99.64\% \\
{\bf  Strings+Wavelet}  &  {\bf 82.97\%}& {\bf 99.65\%}  \\
\hline
\multicolumn{3}{c}{\phantom{abc}} \\
{\bf  Strings+Entropy} &  92.10 \% &  99.97\%  \\
{\bf  Strings+Entropy+Wavelet} & {\bf 94.27 \%} & {\bf 99.98\%}   \\
\multicolumn{3}{c}{\phantom{abc}} \\
{\bf  Strings+Entropy} &  98.63\% &  99.19\%  \\
{\bf  Strings+Entropy+Wavelet} & {\bf 98.90 \%}  & {\bf 99.23\%}  \\
\bottomrule
\end{tabular}
\end{adjustbox}
\caption{ \emph{Wavelet-based decompositions of software entropy boosts performance on a parasitic malware detection task.}   The left hand column shows the hit rate of the model, and the right hand column shows the correct rejection rate. Each pair of rows show the numerical values for points that form approximate vertical slices through the  red and blue ROC curves in Fig. 5. That is, each pair of rows compares hit rates on parasitic malware for approximately equal false positive rates on clean software. }
\label{expt2_results_table}
\end{table*}

\begin{table*}\centering
\ra{1.3}
 \begin{adjustbox}{max width=\textwidth}
\begin{tabular}{@{}lll @{}}\toprule
\toprule
 \phantom{abc} & \multicolumn{2}{c}{{\bf Contribution of Wavelet Features}  \scriptsize \normalsize  }\\
 {\bf Model} &  {\bf \% of All Features} & {\bf \% of Influential Features} \\ \midrule  
{\bf  Strings+Wavelet}   & 2.10\% & {\bf 7.00\%} \\
{\bf  Strings+Entropy+Wavelet}& 1.96\% &  {\bf 4.50\%} \\
\bottomrule
\end{tabular}
\end{adjustbox}
\caption{ \emph{Wavelet-based features are disproportionately likely to be influential features.}  As defined in Section~\ref{expt2_method}, influential features have a particularly strong impact on the machine learning model's classification.}
\label{expt2_importance_table}
\end{table*}

In Figure 5 and  Table~\ref{expt2_results_table}, we compare the performance of the logistic lasso parasitic malware classifier using datasets with and without wavelet features.  In particular, the ROC curves in Fig. 5 graphically depict performance results across a range of decision thresholds, and Table 2 highlights numerical results at particular samples of the ROC curves.  The left hand column of Table 2 shows the hit rate of the model, and the right hand column shows the correct rejection rate. Each pair of rows in Table 2 can be seen as providing concrete values for samples of points from the ROC curves in Fig. 5, where the rows for each pair represent samples from the blue and red curves which have nearly aligned x-coordinates. Thus, each pair of rows describes the effect of adding wavelet features at roughly comparable tolerances for risking a false positive.

The  wavelet features improved the string-only model's ability to detect parasitics while simultaneously reducing false positives.  The effect of wavelet features on detection was fairly strong for most false positive rates.   For example, for false positive rates around one-third of one percent, the wavelet features boosted detection of parasitic malware from 80.90\% to 82.97\% despite only adding $\sim$24k features to the original corpus of $\sim$1.1 million strings.   
Moreover, Fig 5. (right plot) reveals that inclusion of wavelet features boosted the parasitic detection performance of a strings-plus-entropy model in a fairly pronounced way as well.    For false positive rates around .02-.03\%, detection of parasitic malware jumped from 92.10\% to 94.27\%.   For false positive rates around .77-.79\%, detection of parasitic malware jumped from 98.63\% to 98.90\%. These results in Fig. 5 (right side) reinforce the conclusion of Experiment 1, we find that the wavelet features capture information that goes beyond more pedestrian entropy-based information (mean, max, standard deviation, etc.).  Overall, these results suggest that the wavelet energy spectrum extracted from the entropy signal of an executable file provides a useful set of features for a machine learning model for automatically  detecting parasitic malware.   Moreover, the predictive value of these features seems to not be redundant with other, simpler summary features derivable from the entropy signal.


In Table~\ref{expt2_importance_table}, we report some additional results about the most influential features in the various models.  In the strings-only model, we found that the 100 most influential strings in terms of pushing the model towards a parasitics classification included examples such as:   \texttt{\justify{CreateKernelThread, Trampoline, FreeAllBuffers, VVVVVVVVVVVVVVVVVVVVVVVVVV\-VVVVVVVVVVVVVVVVVVVVVVVV, UUUUUUUUUUUUUUUUUUUUUUUUUUUUUUUUUUUUUUUUUUUUUUUU, SetProcessPriorityBoost}}, \texttt{\justify{Cr\-eateProcessA}}, 
and \texttt{\justify{! Best regards 2 Tommy Salo~002E [Nov-2005] yours [Dziad\-ulja Apanas]}}.   For the strings+wavelet model, we see that even though the wavelet features comprise a relatively small proportion (2.1\%) of the strings+wavelet model, they constitute a relatively large proportion (7.0\%) of that model's set of influential features.  From an adversarial point of view, it is a nice finding that wavelet-based features can displace some of the importance of strings, as it is presumably easier for an evasive malware writer to alter a suggestive string such as  \texttt{Trampoline} (the string is suspicious as it evokes derivatives of the state-sponsored Stuxnet parasitic worm) than to displace an entropic energy spectral configuration in a direction favored by a machine learning model.    Finally, in the strings+wavelet+entropy model, wavelet features were also disproportionately influential on the final classification; they were about 2.5 times more likely to be influential features than would have been predicted based on their overall prevalence in the feature corpus alone.

\section{Grand Discussion}

All together, wavelet decompositions on software entropy seem to be useful for malware prediction tasks by capturing the degree to which a portable executable file exhibits suspicious patterns of shifting entropy within its byte-level code.  In particular, we considered the problem that certain kinds of malware (e.g. parasitic malware) tend to contain chunks of encrypted and compressed code embedded in an otherwise normal looking executable file.   To address this situation, we applied a wavelet decomposition to  each file's entropy stream so as to obtain each file's entropic wavelet energy spectrum.  The entropic wavelet energy spectrum characterizes how a file distributes entropic change across multiple levels of spatial resolution.      In the first study, we found that a single feature derived from wavelet decompositions of software entropy can yield valuable predictive information in a heterogeneous corpus of malware.  In the second study, we found that features derived from the wavelet decompositions boosted performance on a large-scale parasitic malware detection task, and that a classifier built solely on three types of features (strings+entropy+wavelet) can produce excellent predictive performance.   In both studies, we found that the information provided by wavelet decompositions of software entropy is not merely redundant with more common measures such as mean entropy or standard deviation of the entropy. 


Future research relating wavelet decompositions to malware classification in machine learning tasks might consider any of the following goals: 
\begin{enumerate}
\item Exploit predictive value from information about the \emph{location} of entropic change (perhaps as pointers for extracting further information about those parts of the file).  This location of entropic change is provided in the mother wavelet coefficients across which we have marginalized to obtain the wavelet energy spectrum. 
\item Apply a more powerful classifier, such as a deep-learning neural network, which could consider more complicated interactions between features when modeling the response.   In addition, incorporate other classes of features (n-grams~\cite{InTheWild}, statistical functions of n-grams~\cite{tabish}, etc.)  What kinds of features interact usefully with the wavelet energy spectrum in predicting malware, and what can we learn from that about the existing corpus of parasitic malware?\footnote{Note that the predictive performance of the model would likely improve by first applying appropriate dimensionality reduction techniques; see e.g.~\cite{wojnowicz3}.}
\item Investigate the potential utility of \emph{non-entropic} wavelet energy spectra from byte-level representations of executable files.  Indeed, entropy streams are just one possible example of real-valued streams derivable from byte-level file content (see e.g.~\cite{tabish}), and wavelet energy spectra can be extracted from \emph{any} real-valued function on the raw bytes.  
\end{enumerate}

\end{document}